\documentclass[journal]{IEEEtran}

\usepackage{mathtools}
\usepackage{amssymb}
\newcommand\mybar{\kern1pt\rule[-\dp\strutbox]{1pt}{\baselineskip}\kern1pt}
\usepackage{pifont}% http://ctan.org/pkg/pifont
\newcommand{\cmark}{\ding{51}}%
\newcommand{\xmark}{\ding{55}}%

\usepackage{dsfont}

\newcommand{\R}{\mathbb{R}}
\usepackage{cite}
\usepackage{amsmath,amssymb,amsfonts}
\usepackage{amsmath}
\usepackage{mathrsfs}
\usepackage{graphicx}
\usepackage{textcomp}
\usepackage[table,xcdraw]{xcolor}
\usepackage{ragged2e}
\usepackage{bbm}
\usepackage{multirow}
\usepackage[linesnumbered, ruled]{algorithm2e}
\usepackage{algpseudocode}
\usepackage[caption=false, font=footnotesize]{subfig}
\usepackage{mathtools} 
\usepackage[hang,flushmargin]{footmisc}
\usepackage{lipsum}
\makeatletter
\newcommand{\algorithmfootnote}[2][\footnotesize]{%
  \let\old@algocf@finish\@algocf@finish% Store algorithm finish macro
  \def\@algocf@finish{\old@algocf@finish% Update finish macro to insert "footnote"
    \leavevmode\rlap{\begin{minipage}{\linewidth}
    #1#2
    \end{minipage}}%
  }%
}

\makeatletter
\def\ps@IEEEtitlepagestyle{%
  \def\@oddfoot{\mycopyrightnotice}%
  \def\@oddhead{\hbox{}\@IEEEheaderstyle\leftmark\hfil\thepage}\relax
  \def\@evenhead{\@IEEEheaderstyle\thepage\hfil\leftmark\hbox{}}\relax
  \def\@evenfoot{}%
}
\def\mycopyrightnotice{%
  \begin{minipage}{\textwidth}
  \scriptsize
   1551-3203 \textcopyright 2020 IEEE. Personal use of this material is permitted.
  Permission from IEEE must be obtained for all other uses, in any current or future
  media, including reprinting/republishing this material for advertising or promotional
  purposes, creating new collective works, for resale or redistribution to servers or
  lists, or reuse of any copyrighted component of this work in other works.
  DOI: \href{https://doi.org/10.1109/TII.2020.3048391}{ 10.1109/TII.2020.3048391}.
  \end{minipage}
}
\makeatother

\usepackage[flushleft]{threeparttable}
\usepackage{etoolbox}
\appto\TPTnoteSettings{\footnotesize}

\usepackage{hyperref}
\hypersetup{
    colorlinks=true,
    linkcolor=blue,
    filecolor=magenta,      
    urlcolor=cyan,
}

\newcommand{\thickhline}{%
    \noalign {\ifnum 0=`}\fi \hrule height 1pt
    \futurelet \reserved@a \@xhline
}
\newcolumntype{"}{@{\hskip\tabcolsep\vrule width 1pt\hskip\tabcolsep}}

\usepackage{tabu}

\begin{document}

\title{CovTANet: A Hybrid Tri-level Attention Based Network for Lesion Segmentation, Diagnosis, and Severity Prediction of COVID-19 Chest CT Scans}

\author{Tanvir~Mahmud,
        Md.~Jahin~Alam,
        Sakib~Chowdhury,
        Shams~Nafisa~Ali,
        Md~Maisoon~Rahman,\\
        ~Shaikh~Anowarul~Fattah,~\IEEEmembership{Senior~Member, IEEE,}
        and~Mohammad~Saquib,~\IEEEmembership{Senior~Member, IEEE}% <-this % stops a space
\thanks{T.~Mahmud, M.~J.~Alam, S.~Chowdhury, M.~M.~Rahman, and S.~A.~Fattah are with the Department
of Electrical and Electronic Engineering, and S.~N.~Ali is with the Department of Biomedical Engineering, Bangladesh University of Engineering and Technology, Bangladesh, e-mail: \{tanvirmahmud, fattah\}@eee.buet.ac.bd, \{jahinalam.eee.buet, sakibchowdhury131, snafisa.bme.buet, 2maisoon1998\}@gmail.com}% <-this % stops a space
\thanks{M~Saquib is with Department of Electrical Engineering, The University of Texas at Dallas, USA,
email: saquib@utdallas.edu
}}% <-this % stops 

\markboth{IEEE Transactions on Industrial Informatics}%
{Tanvir \MakeLowercase{\textit{et al.}}: CovTANet: A Hybrid Tri-level Attention Based Network}

\maketitle

%\copyrightnotice

\begin{abstract}
Rapid and precise diagnosis of COVID-19 is one of the major challenges faced by the global community to control the spread of this overgrowing pandemic. In this paper, a hybrid neural network is proposed, named CovTANet, to provide an end-to-end clinical diagnostic tool for early diagnosis, lesion segmentation, and severity prediction of COVID-19 utilizing chest computer tomography (CT) scans. A multi-phase optimization strategy is introduced for solving the challenges of complicated diagnosis at a very early stage of infection, where an efficient lesion segmentation network is optimized initially which is later integrated into a joint optimization framework for the diagnosis and severity prediction tasks providing feature enhancement of the infected regions. Moreover, for overcoming the challenges with diffused, blurred, and varying shaped edges of COVID lesions with novel and diverse characteristics, a novel segmentation network is introduced, namely Tri-level Attention-based Segmentation Network (TA-SegNet). This network has significantly reduced semantic gaps in subsequent encoding decoding stages, with immense parallelization of multi-scale features for faster convergence providing considerable performance improvement over traditional networks. Furthermore, a novel tri-level attention mechanism has been introduced, which is repeatedly utilized over the network, combining channel, spatial, and pixel attention schemes for faster and efficient generalization of contextual information embedded in the feature map through feature re-calibration and enhancement operations. Outstanding performances have been achieved in all three-tasks through extensive experimentation on a large publicly available dataset containing 1110 chest CT-volumes that signifies the effectiveness of the proposed scheme at the current stage of the pandemic.
\end{abstract}


\IEEEpeerreviewmaketitle

\section{Introduction}

\IEEEPARstart{S}{ince} the onset of the coronavirus disease (COVID-19) in December-2019, it has severely jeopardized the global healthcare systems for its extremely infectious nature. 
With its extreme infectious nature and high mortality rate, it has been declared as one of the most devastating global pandemics of history~\cite{b1}. 
Although reverse transcription-polymerase chain reaction (RT-PCR) assay is considered as the gold standard for COVID-19 diagnosis, the shortage of this expensive test-kit coupled with the elongated testing protocol and relatively low sensitivity (60-70\%) calls for an alternative diagnostic tool that is adequately efficient to perform prompt mass-screening~\cite{rt}. For providing immediate and proper clinical support to the critical patients, severity quantification of the infection is also a dire need. With numerous success stories in the field of clinical diagnostics and biomedical engineering, artificial intelligent (AI) assisted diagnostic paradigms can be embraced as a medium of paramount importance to conduct automated diagnosis and severity quantification of COVID-19 with substantial accuracy and efficiency~\cite{ai, abdel3}.
Several deep learning-based frameworks have been explored in recent times deploying automated screening of chest radiography and computer tomography as one of the vital sources of information for COVID diagnosis~\cite{ct1,ct2,abdel2, x1, abdel}.  However, owing to the relatively higher sensitivity and the provision of enhanced infection visualization in the three-dimensional representation, CT-based screening is a more viable alternative than the X-ray counterparts. 
With a large number of asymptomatic patients, early detection of COVID-19 through imaging modalities is still a stupendously challenging task due to significantly smaller, scattered, and obscure regions of infections that are difficult to distinguish~\cite{early}. These diverse heterogeneous characteristics of infections among different subjects also make the severity prediction to be an extremely difficult objective to achieve~\cite{severity}. The scarcity of considerably large reliable datasets further increases the complexity of the endeavor. Recent studies mostly opt for solving this daunting task partially where infection segmentation, diagnosis, or severity analysis have separately attempted~\cite{inf, mini, cop}. Such methods lack the complete integration of the objectives for providing a robust clinical tool.

\begin{figure*}[t]
    \centering
%    \centerline{\includegraphics[width=0.95\textwidth]{GA.jpg}}
    \includegraphics[scale=0.445]{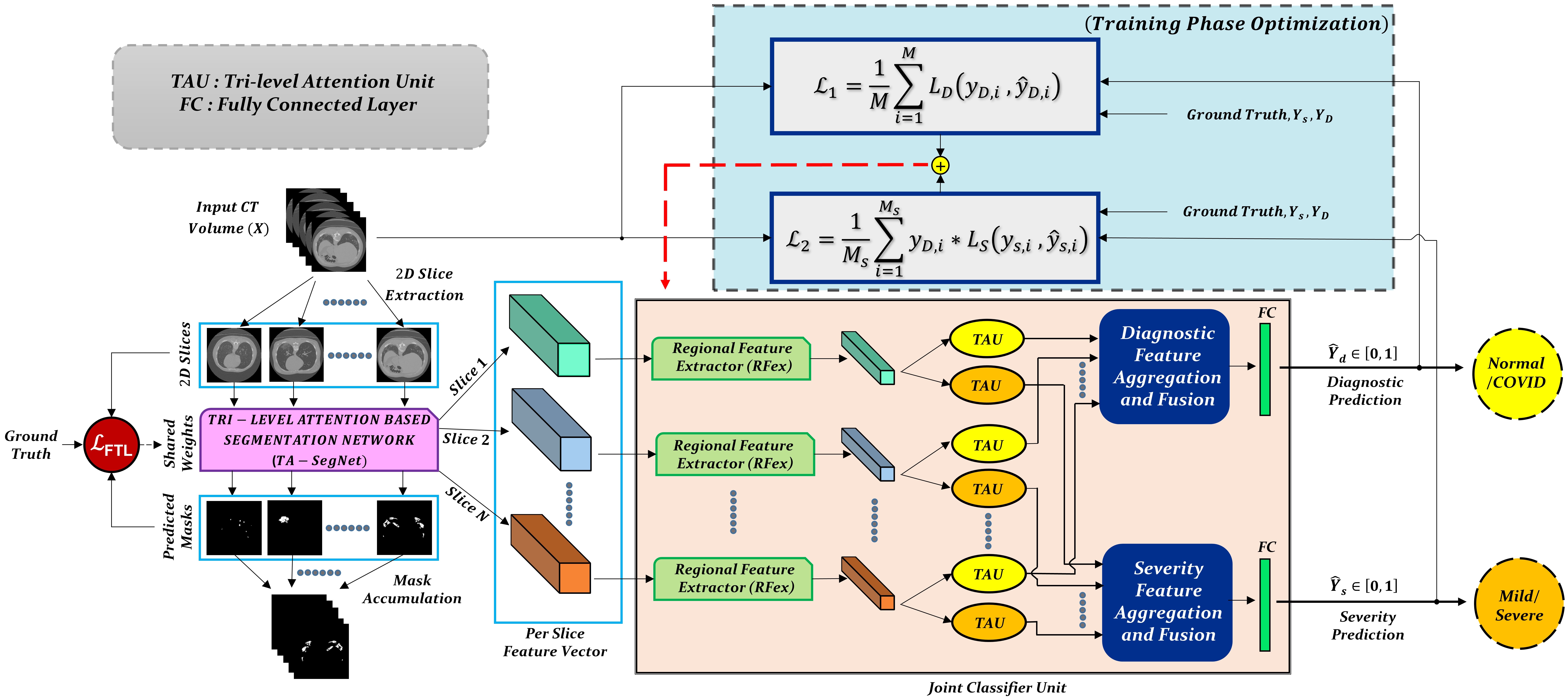}
    \caption{\textbf{Graphical overview of the optimization scheme of CovTANet: Tri-level Attention based Segmentation Network (TA-SegNet) extracts the slicewise lesion segmentation mask and representational features of the corresponding CT-volume which are employed later for the joint optimization of severity prediction and diagnosis. Separate Tri-level Attention Units (TAUs) are employed to enhance the diagnostic features and severity based features in the joint optimization process.}}
    \label{f1}
\end{figure*}

Deep learning-based approaches have been widely incorporated in diverse medical imaging applications for their unprecedented performance even in challenging conditions.  Several state-of-the-art networks have been investigated for the segmentation of COVID lesions from chest CTs including FCN~\cite{new2}, U-Net~\cite{unet}, UNet++~\cite{unet++}, and ResUNet~\cite{ct2}. However, precise segmentation of COVID lesions has still been a major challenge due to the patchy, diffused, and scattered distributions of the infections involving ground-glass opacities, pleural effusions, and consolidations~\cite{new1}. Traditional U-Net and its variants with similar encoder-decoder architectures suffer from increased semantic gaps between the corresponding scale of feature maps of encoder-decoder modules while experiencing vanishing gradient problems with several optimization issues due to the sequential optimization strategy of multi-scale features. Moreover, the contextual information generated at different scales of representation does not properly converge into the final reconstruction of the segmentation mask that results in sub-optimal performances. Lately, numerous attention-gated mechanisms have been paired up with the traditional segmentation frameworks and demonstrated highly promising performance in terms of gathering more contextual information through redistribution of the feature space~\cite{attention, inf}.

\begin{algorithm}[!t]
\DontPrintSemicolon
\SetAlgoLined
\SetNoFillComment
\LinesNotNumbered 
%\SetSideCommentLeft
%\tcc{iterate over all training examples}
 \KwData{CT-Volume Data, $\mathbf{V}$; Segmentation masks, $\mathbf{M}$; Diagnostic labels, $\mathbf{Y_D}$; Severity labels, $\mathbf{Y_S}$}

 \KwResult{weight matrices, $W_{TA-SegNet}$, $W_{RF_{ex}}$, $W_{TAU}$, $W_{F_d}$, $W_{F_s}$, $W_{fc_1}$, $W_{fc_2}$}
 
 \tcc{Optimize the TA-SegNet}
  \tcp*[l]{Extract 2D CT-slices and masks}
    Get $\mathbf{S_V}$ = Slice-Extractor $(\mathbf{V})$ \;
    Get $\mathbf{S_M}$ = Slice-Extractor $(\mathbf{M})$ \;
        
    Initialize weights $W_{TA-SegNet}$\;
     \While{training loss, $L_{Seg} >$ threshold ($\epsilon$)}{
        Calculate, $\mathbf{S_{M}^p} = W_{TA-SegNet}(\mathbf{S_V})$ \;  
        Calculate loss, $L_{Seg} = \mathscr{L}_{FTL}\mathbf{(S_{M}^p, S_{M})}$ \;
        Run optimizer and update $W_{TA-SegNet}$ \;
    }
  
 \tcc{Optimize the Classifier Unit}
  Initialize $W_{RF_{ex}}$, $W_{TAU}$, $W_{F_d}$, $W_{F_s}$, $W_{fc_1}$, $W_{fc_2}$ \;
  
  \While {training loss, $\mathscr{L} >$ threshold ($\epsilon$)}{
    \For{$i\leftarrow 1$ \KwTo $N$}{
        \tcp*[l]{Extract Per-Slice Features}
        \For{$j\leftarrow 1$ \KwTo $s$}{
            Get ${f_{i,j}} = W_{TA-SegNet}^{intmd} (S_{V_i})$ \;
            Get ${F_{i,j}} = W_{RF_{ex}} (f_{i,j})$ \;
        }
    \tcp*[l]{Aggregate volumetric features}
    
    Get $A_{diagnostic} = W_{F_d}(W_{TAU}(F_{i,k}))\mid_{k=1}^{k=s}$ \;
    Get $A_{severity} = W_{F_s}(W_{TAU}(F_{i,k}))\mid_{k=1}^{k=s}$ \;
    
    \tcp*[l]{Generate predictions}
    Get $Y_D^p$ = $ \sigma( W_{fc_1} (A_{diagnostic}))$ \;
    Get $Y_S^p$ = $\sigma (W_{fc_2} (A_{severity}))$ \;
    }
    
    Calculate $\mathscr{L} = \mathcal{L}_1(\mathbf{Y_D, Y_D^p}) + \mathcal{L}_2(\mathbf{Y_D, Y_S, Y_S^p})$ \;
    Run optimizer and update $W_{RF_{ex}}$, $W_{TAU}$, $W_{F_d}$, $W_{F_s}$, $W_{fc_1}$, $W_{fc_2}$\;
  }
  \algorithmfootnote{$N, s$ denote number of CT-volumes and slices per-volume, respectively.\\
  $W_{TA-SegNet}^{intmd}$ represents the intermediate part of the TA-SegNet weight matrix for providing the per-slice feature vector. 
  }
   \label{alg}
 \caption{Training and Optimization of CovTANet}
\end{algorithm}

In this paper, CovTANet, an end-to-end hybrid neural network is proposed, that is capable of performing precise segmentation of COVID lesions along with accurate diagnosis and severity predictions. The intricate network of the proposed scheme emerges as an effective solution by overcoming the limitations of the traditional approaches.  The major contributions of this work can be summarized as follows:
\begin{enumerate}
    \item A novel tri-level attention guiding mechanism is proposed combining channel, spatial and pixel domains for feature recalibration and better generalization.

    \item A tri-level attention based segmentation network (TA-SegNet) network is proposed for precise segmentation of COVID lesions integrating the triple attention mechanisms with parallel multi-scale feature optimization and fusion.
    
    \item A multi-phase optimization scheme is introduced by effectively integrating the initially optimized TA-SegNet with the joint diagnosis and severity prediction framework. 

    \item A system of networks is proposed for efficient processing of CT-volumes to integrate all three objectives for improving performance in challenging conditions.
    
    \item Extensive experimentations have been carried out over a large number of subjects with diverse levels and characteristics of infections.
    
\end{enumerate}

%Provision of enhanced ground glass opacity (GGO) and three-dimensional representation of lung consolidations and bilateral, mutifocal patchy shadows in CT images coupled with a relatively high sensitivity make CT images a vital complementary tool compared to X-rays in screening COVID-19 [12].

%Segmentation is very tough

%Diagnosis is impossible

%We have solved everything.

\section{Methodology}
\begin{figure}[t]
    \centering
    \includegraphics[scale=0.13]{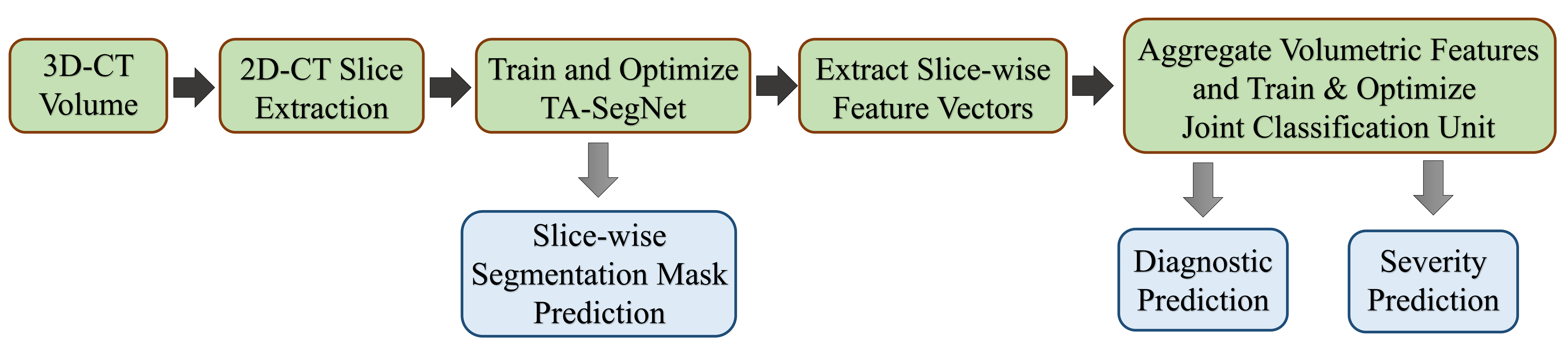}
    \caption{\textbf{Optimization flowchart of the proposed CovTANet network.}}
    \label{flow}
\end{figure}

The proposed CovTANet network is developed in a modular way focusing on diverse clinical perspectives including precise COVID diagnosis, automated lesion segmentation, and effective severity prediction. The whole scheme is represented in Fig.~\ref{f1}. Here, 
a hybrid neural network (CovTANet) is introduced for segmenting COVID lesions from CT-slices as well as for providing effective features of the region-of-lesions which are later integrated for the precise diagnosis and severity prediction tasks.
The complete optimization process is divided into two sequential stages for efficient processing. Firstly, a neural network, named as Tri-level Attention-based Segmentation Network (TA-SegNet), is designed and optimized for slicewise lesion segmentation from a particular CT-volume. A tri-level attention gating mechanism is introduced in this network with multifarious architectural renovations to overcome the limitations of the traditional Unet network (Section~\ref{lim}), which gradually accumulates effective features for precise segmentation of COVID lesions. 
Because of the pertaining complicacy with blurred, diffused, and scattered patterns of COVID lesions, it is quite obvious that direct utilization of the final segmented portions for diagnosis may result in loss of information due to some false positive estimations. 
The proposed CovTANet aims to resolve this issue by extracting effective features regarding the regions-of-infection utilizing the initially optimized TA-SegNet as it is optimized for precisely segmenting COVID lesions with diverse levels, types, and characteristics.
%segmenting COVID lesions with diverse levels, types, and characteristics and extracting potent features from each of the infected regions. 
%Nevertheless, the proposed TA-Unet is capable of extracting effective features regarding the regions-of-infection as it is capable of segmenting COVID lesions with diverse levels, types, and characteristics with unprecedented precision.
Therefore, slicewise effective features are extracted utilizing the optimized TA-SegNet network and deployed into the second phase of training for the joint optimization of diagnosis and severity prediction tasks. Additionally, separate regional feature extractors are employed for generating more generalized forms of the slicewise feature vectors from different lung regions. Subsequently, these generalized feature representations of CT-slices are guided into separate volumetric feature aggregation and fusion schemes through the proposed tri-level attention mechanism for extracting the significant diagnostic features as well as severity based features. The diagnostic path is supposed to extract the more generalized representation of infections while the severity path is more concerned with the levels of infections. Both the diagnostic and severity predictions are optimized through a joint optimization strategy with an amalgamated loss function. The whole training and optimization process is summarized in Algorithm~\ref{alg}. In addition, the optimization flow of the complete CovTANet network is shown in Fig.~\ref{flow}. Several architectural submodules of the CovTANet are discussed in detail in the following sections.

\subsection{Proposed Tri-level Attention Scheme}
Attention mechanism, first proposed in~\cite{attention2} for enhanced contextual information extraction in natural language processing, has been adopted in numerous fields including medical image processing~\cite{squeeze, attention3}. 
This mechanism assists faster convergence with considerable performance improvement by eliminating the redundant parts while putting more attention on the region-of-interests through the generalization of the predominant contextual information.
%Since the traditional convolutional filtering process relies on the local observation window and thus, it gets complicated to generalize the relevant features through a deep stack of convolutional and pooling layers. The attention mechanism helps to reorganize the feature space through the generalization of the predominant contextual information to enhance the relevant features.
In this work, we have proposed a novel self-supervised attention mechanism combining three levels of abstraction for improved generalization of the relevant contextual features, i.e. channel-level, spatial-level, and pixel-level. The channel attention (CA) mechanism operates on a broader perspective to emphasize the corresponding channels containing more information, while the spatial attention (SA) mechanism concentrates more on the local spatial regions containing region of interests, and finally, the pixel attention (PA) mechanism operates on the lowest level to analyze the feature relevance of each pixel. However, relying only on the higher level of attention causes loss of information while relying on lower/local levels may weaken the effect of generalization. Hence, to reach the optimum point of generalization and re-calibration of feature space, we have introduced a tri-level attention unit (TAU) mechanism that integrates the advantages of all three levels of attention. This TAU unit module is repeatedly used all over the CovTANet network (Fig.~\ref{f1}) to improve the feature relevance through feature recalibration.
%Contributions

\begin{figure*}[t]
    \centering
    \includegraphics[scale=0.095]{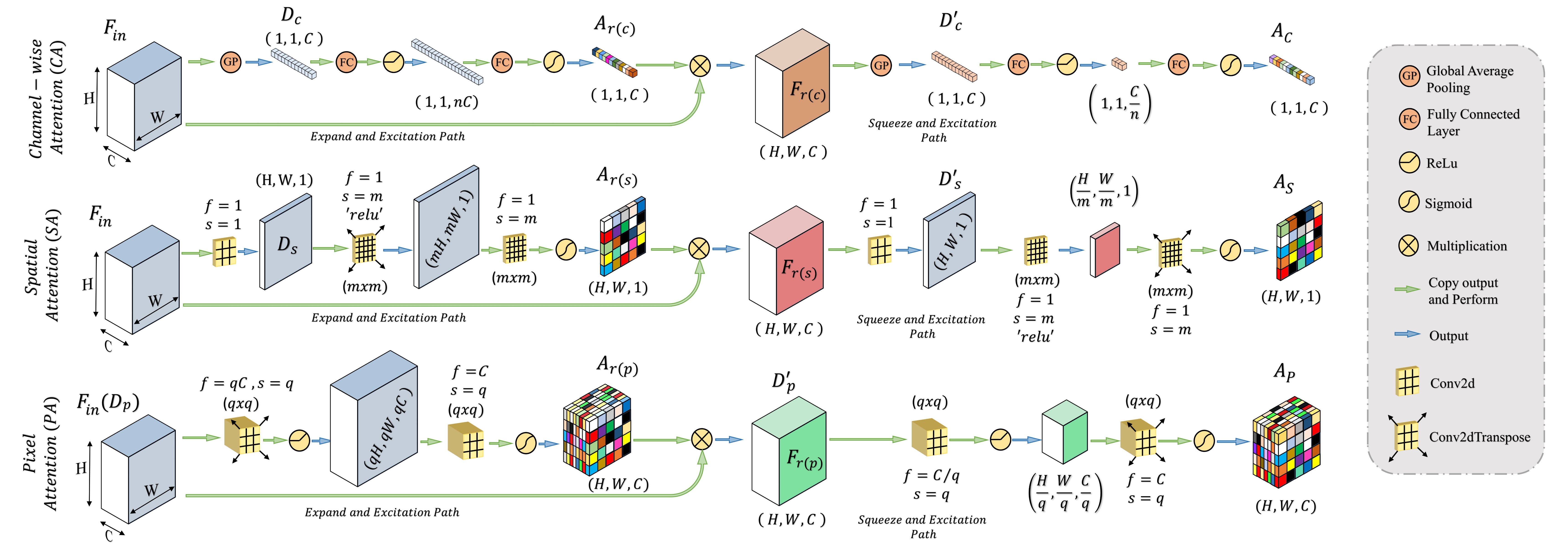}
    \caption{\textbf{Schematic of the proposed channel, spatial, and pixel attention mechanisms. Each attention mechanism integrates feature re-calibration operation through expand-excitation scheme followed by feature generalization operation through squeeze-excitation scheme.}}
    \label{f2}
\end{figure*}

\begin{figure}[t]
    \centering
    \includegraphics[width=\columnwidth]{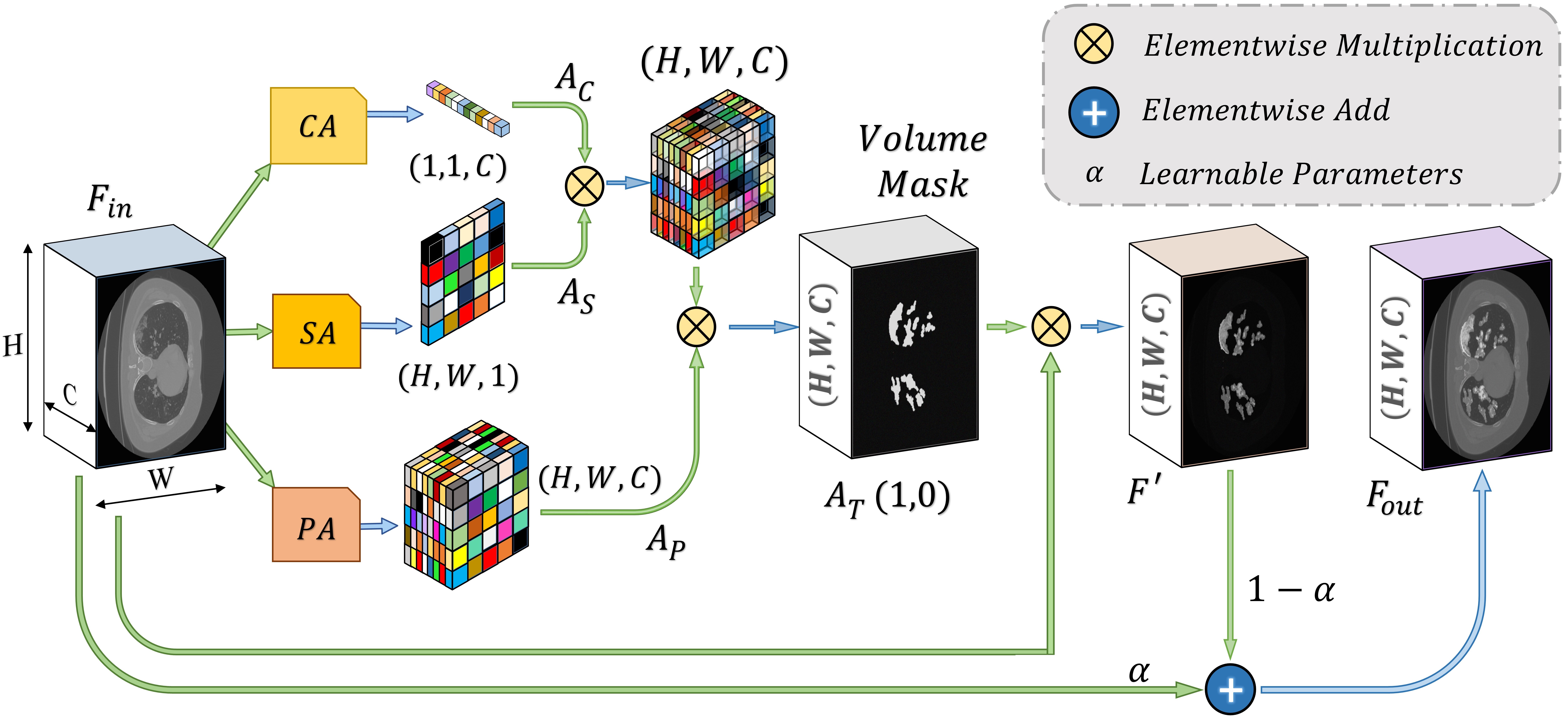}
    \caption{\textbf{Schematic of the proposed Tri-level Attention Unit (TAU) integrating channel attention (CA), spatial attention (SA), and pixel attention (PA) mechanisms. Here, channel broadcasting operation is carried out before element wise multiplication/addition of feature maps.}}
    \label{f3}
\end{figure}

In general, the proposed attention mechanisms operating at different levels of abstraction (shown in Fig.~\ref{f2}) can be divided into two phases: a feature re-calibration phase followed by a feature generalization phase. In each phase, a statistical description of the intended level of generalization is extracted, which is processed later for generating the corresponding attention map. Let, $F_{in}\in \R^{H\times W \times C}$ be the input feature map where $(H, W, C)$ represent the height, width, and channels of the feature map, respectively. Here, channel description, $D_{c}\in \R^{1\times1\times C}$ is generated by taking the global averages of the pixels of particular channels, while the spatial description, $D_{s}\in \R^{H\times W \times 1}$ is created by convolutional filtering, and the input feature map, $F_{in}$ represents the pixel description, $D_p \in \R^{H\times W \times C}$ itself.

\begin{figure*}[t]
    \centering
    \includegraphics[scale=0.42]{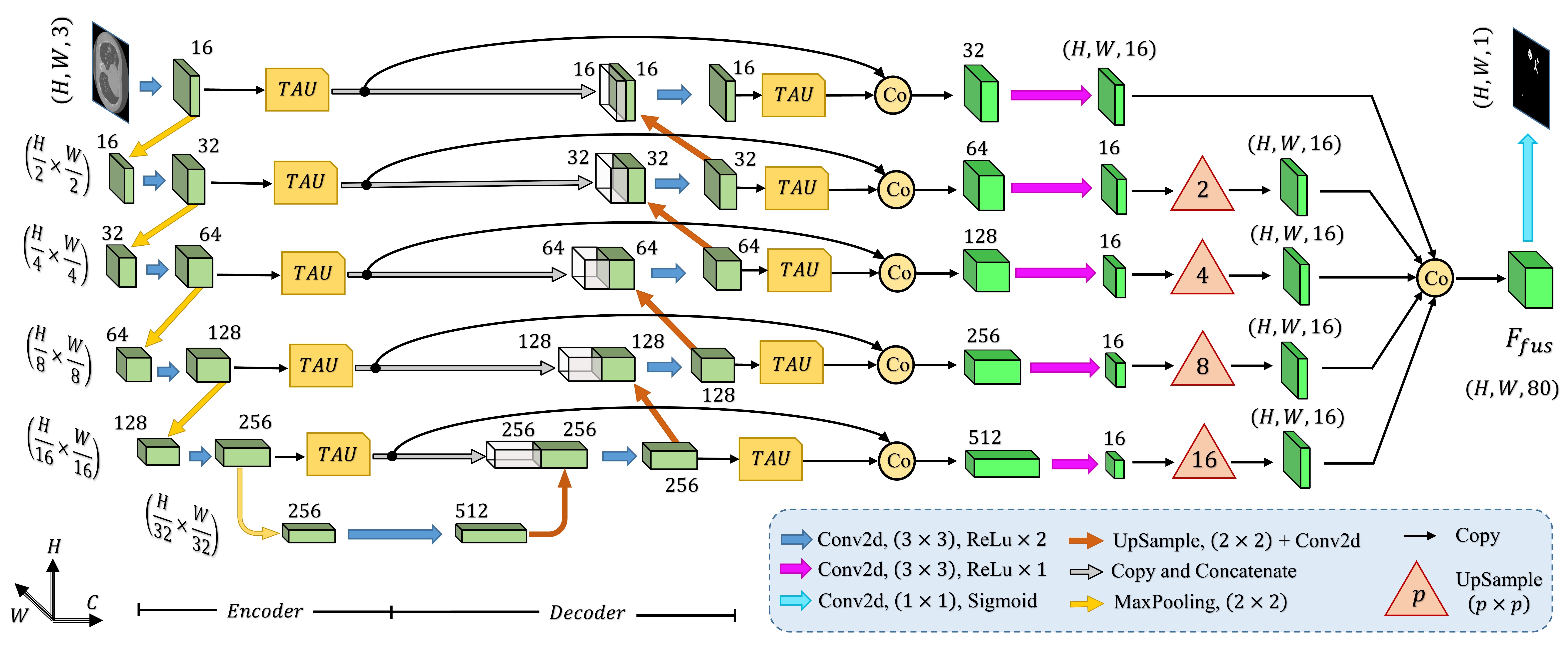}
    \caption{\textbf{Schematic representation of the proposed Tri-level Attention-based Segmentation Network (TA-SegNet) integrating numerous tri-level attention unit (TAU) modules for semantic gap reduction between encoder and decoder modules as well as for efficient reconstruction of lesion-mask.}}
    \label{f4}
\end{figure*}

\begin{figure}[t]
    \centering
    \includegraphics[scale=0.42]{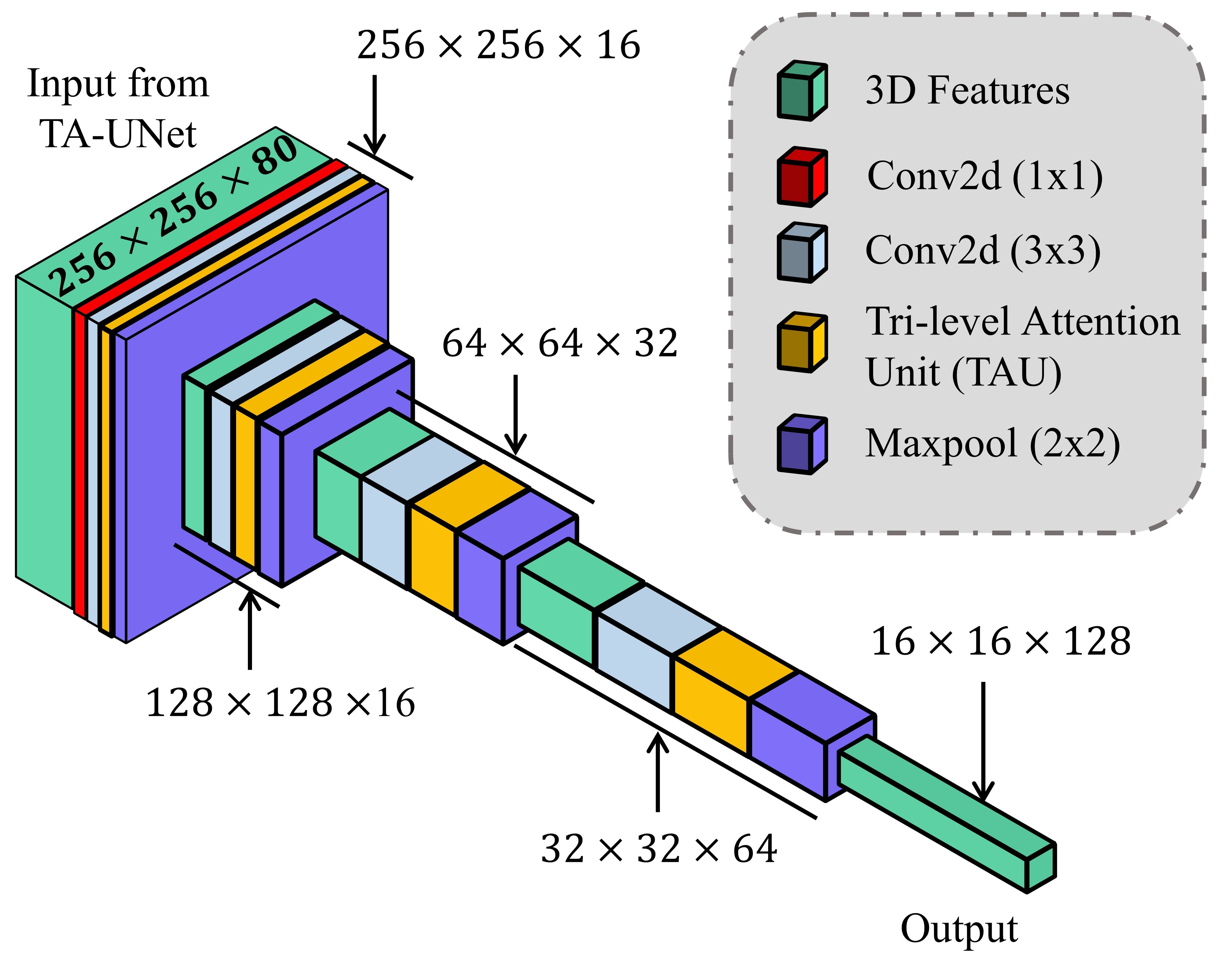}
    \caption{\textbf{Representation of the proposed regional feature extractor module}}
    \label{f5}
\end{figure}

Afterwards, the feature re-calibration phase is carried out by projecting the descriptor vector $D$ to a higher dimensional space followed by the restoration process of the original dimension to generate the re-calibration attention map $A_r$, which is utilized to obtain the re-calibrated feature map $F_r$. This process assists in the redistribution of the feature space in the subsequent feature generalization phase for better generalization of features through sharpening the effective representative features. It can be represented as:
\begin{align}
    F_r = F_{in} \otimes A_r &= F_{in} \otimes \sigma(W_R(W_E(D))) \\
    &= F_{in} \otimes \sigma(W_R(W_E(W_D(F_{in}))))
\end{align}
where $\otimes$ represents the element-wise multiplication with the required dimensional broadcasting operation, $W_D$ denotes the statistical descriptor extractor, $W_E$ represents the dimension expansion filtering, $W_R$ represents the dimension restoration filtering, and $\sigma(\cdot)$ represents the sigmoid activation. For the channel-attention mechanism, $W_E$ and $W_R$ are realized by fully connected layers, while for spatial and pixel attention, convolutional filters are employed.

Subsequently, the feature generalization operation is carried out through the squeeze and excitation operation on the re-calibrated feature space, $F_r$ to generate the 
effective attention map $A$. In this phase, the extracted feature descriptor, $D'$ is projected into a lower-dimensional space to extract the most effective representational features and thereafter, reconstructed back to the original dimension. Such sequential dimension reduction and reconstruction operations provide an opportunity to emphasize the generalized features while reducing the redundant features. Hence, the generated attention map $A$ provides the opportunity to reduce the effect of redundant features by providing more attention to the effective features, and it can be represented as:  
\begin{equation}
    A = \sigma(W_{R'}(W_S(D'))) = \sigma(W_{R'}(W_S(W_{D'}(F_r)))) 
\end{equation}
where $W_S, W_{R'}$ represents the corresponding squeeze and restoration filtering, respectively, while $W_{D'}$ represents the statistical descriptor extractor. Therefore, three levels of attention maps are generated, i.e. a channel attention map $A_C \in \R^{1\times 1\times C}$, a spatial attention map $A_S \in \R^{H\times W \times 1}$, and a pixel attention map $A_P \in \R^{H\times W\times C}$. The tri-level attention unit (TAU), represented in Fig.~\ref{f3}, generates the effective volumetric, triple attention mask $A_T$ integrating all three maps, which is given by:
\begin{equation}
    A_T = A_P \otimes (A_S \otimes A_C)
\end{equation}
%where $\otimes$ represents element-wise multiplication with necessary dimensional broadcasting.

Later, this accumulated attention mask $A_T$ is used to transform the input feature map $F_{in}$ to $F'$ for enhancing the region-of-interest, and finally the output feature map, $F_{out}$ is generated through the weighted addition of the input and transformed feature maps, and these can be summarized as:
\begin{align}
    F' = F_{in} \otimes A_T \\
    F_{out} = T(F_{in}) = \alpha F_{in} + (1-\alpha) F'
\end{align}
where T($\cdot$) represents the proposed Tri-level attention mechanism, $\alpha$ is a learnable parameter that is optimized through the back-propagation algorithm along with other parameters.

\subsection{Proposed Tri-level Attention-based Segmentation Network (TA-SegNet)}
\label{lim}
The proposed TA-SegNet network is deployed for segmenting the infected lesions as well as for extracting features for the following joint diagnosis and segmentation tasks (as shown in Fig.~\ref{f1}). For better segmentation, this network introduces several modifications over traditional networks which are mostly based on Fully convolutional networks (FCN) and Unet networks generally.

FCN and Unet are the most widely explored networks for medical image segmentation. 
In FCN, a single stage of encoder module is employed to generate different scales of encoded feature maps from the input image, and afterwards, the segmentation mask is reconstructed through joint processing of multi-scale encoded features.
%In FCN, an encoder module is employed to generate different scales of encoded feature maps from the input image where the feature map is gradually downscaled to produce more generalized representations. Afterwards, these multi-scale feature representations are made uniform through upsampling, aggregated, and processed together for generating the segmentation mask. 
Whereas, the Unet network considerably improved the performance by introducing a decoder module followed by the encoder module to sequentially gather the contextual information of the segmentation mask. Moreover, to recover the loss of information through downscaling, each level of encoder and decoder modules are directly connected through skip connections in Unet. 
Despite that, a semantic gap is generated through such direct skip connections between the corresponding scale of feature maps of the encoder and decoder modules, which hinders proper optimization. For the deeper implementation of the encoder/decoder module, this network further suffers from vanishing gradient problem since different scales of feature maps are optimized sequentially. 

The proposed TA-SegNet network (shown in Fig.~\ref{f4}) integrates the advantages of both Unet and FCN by introducing an encoder-decoder based network with reduced semantic gaps along with the opportunity of parallel optimization of multi-scale features. Firstly, the input images pass through sequential encoding stages with convolutional filtering followed by sequential decoding operations similar to the Unet. Moreover, the output feature map generated from each layer of the encoder unit is connected to the corresponding decoder layer through a Tri-level Attention Unit (TAU) mechanism for better reconstruction in the decoder unit. 
For further generalization and refinement of contextual features, all scales of decoded feature representations also pass through another stage of the attention mechanism. Afterwards, for introducing joint optimization of multi-scale features, the attention gated, refined feature maps generated at different stages of encoder and decoder modules are accumulated through a series of operational stages. Initially, sequential concatenation of corresponding encoder-decoder layer outputs (after attention-gating) are carried out. Following that, channel downscaling operations through convolutional filtering and bi-linear spatial upsampling operations are employed to produce feature vectors with uniform dimensions. Afterwards, these uniform feature vectors are accumulated through channel-wise concatenation to generate the fusion vector $F_{fus}$, and it can be represented as: 
\begin{equation}
    F_{fus} = \mathcal{F}_{i=1}^{N}(T(E_i) \oplus T(D_i))\ %\forall i=\{1, 2,\dots, N\}
\end{equation}
where $\oplus$ represents feature concatenation, $E_i, D_i$ stand for $i_{th}$ level of feature representations from total $N$ levels of the encoder, and decoder modules, respectively, $T(\cdot)$ represents the tri-level attention unit operation, and $\mathcal{F}(\cdot)$ represents the multi-scale feature fusion operation.

Afterwards, the final convolutional filtering is operated on the fusion feature map ($F_{fus}$) to produce the output segmentation mask. Moreover, to introduce transfer-learning in this TA-SegNet similar to other networks, the encoder module can be replaced by different pre-trained backbone networks for better optimization. 
Hence, the proposed TA-SegNet facilitates faster convergence through parallel optimization of the multi-scale features while effectively extracting the region-of-interest from each scale of representation with the novel tri-level attention gating mechanism for providing the optimum performance even in the most challenging conditions.

\subsection{Proposed Regional Feature Extractor Module}
Though the proposed TA-SegNet is optimized for providing precise segmentation performances on challenging COVID lesion extraction, some loss of information is expected to occur especially at the early stages of infection when it is difficult to extract relatively smaller and scattered infection patches. To overcome the limitation, the final fusion vector $F_{fus}$ generated at TA-SegNet is incorporated into further processing, instead of the segmented lesion, as it contains the effective feature representations of the region-of-infections. For further emphasizing the COVID lesion features, a regional feature extractor module ($RF_{ex}$) is also proposed that separately operates on each of the slice-wise fusion vector $F_{fus}$ and thus generates the effective regional feature representation $F_{reg}$. From Fig.~\ref{f1}, it is to be noted that such regional feature extractor module separately operates on the extracted feature vectors of each CT-slice and hence, enhance the effective regional features regarding the infection. 
The architectural details of this module are presented in Fig.~\ref{f5}. It consists of several stages of convolutional filtering while incorporating the Tri-level Attention Unit at each stage. 
These attention units operated at different stages are supposed to execute different roles.
As we go deeper into this $RF_{ex}$ module, more generalized feature representations are created through subsequent pooling operations where the information is made more sparsely distributed among increased channels. Hence, the attention units at earlier stages enhance the more detailed, localized feature representations, while at deeper stages the attention units learn to expedite the generalization process. Therefore, the regional feature extractor module effectively incorporates the proposed tri-level attention mechanism to extract the most generalized representative features of infections from different regions of the respective CT volume.

\subsection{Volumetric Feature Aggregation and Fusion Module}
\begin{figure}[t]
    \centering
    \includegraphics[scale=0.4]{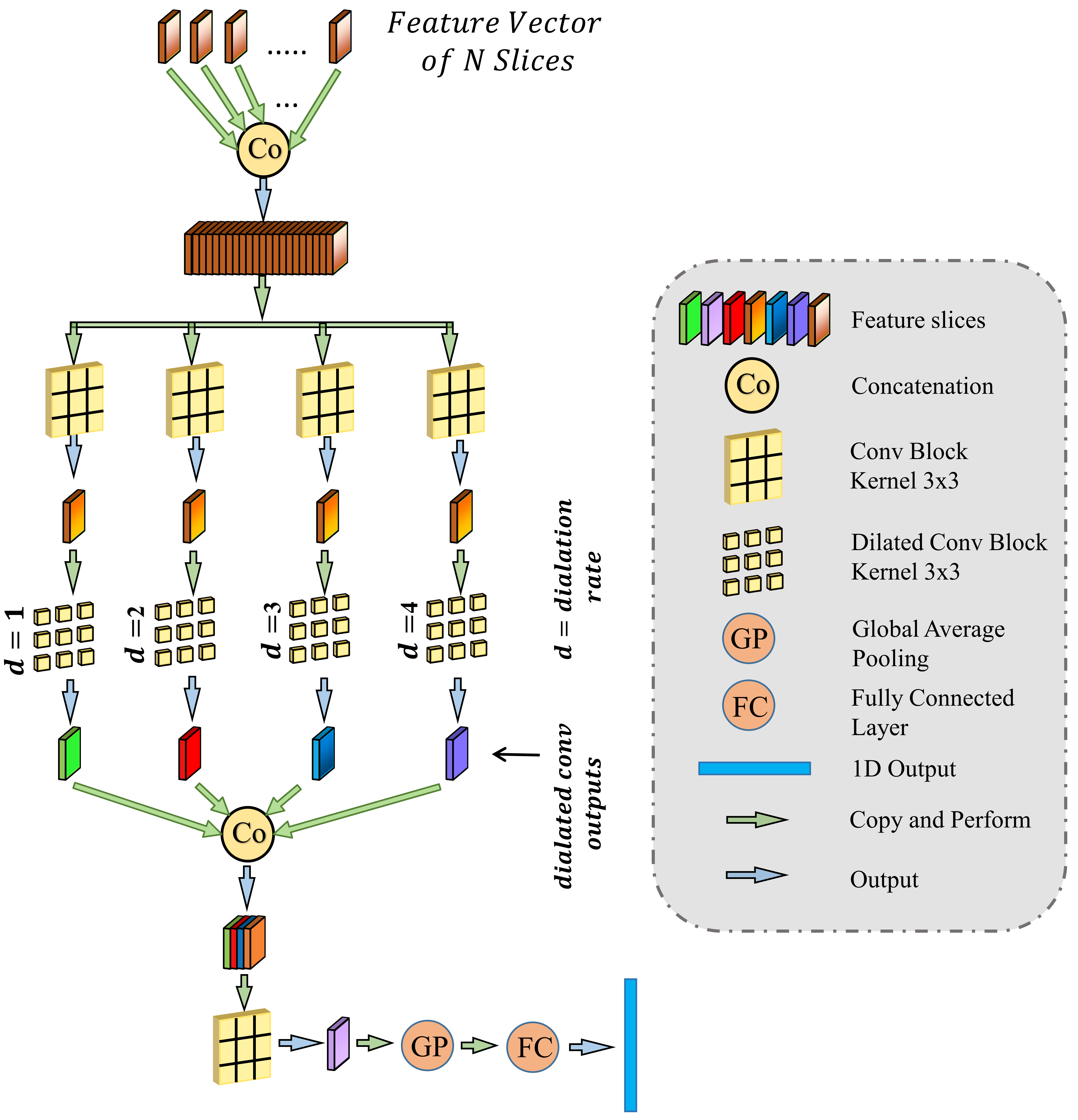}
    \caption{\textbf{Proposed volumetric feature accumulation and fusion scheme used for severity and diagnostic feature extraction}}
    \label{f6}
\end{figure}

The regional features extracted from each slice of the CT volume are supposed to optimize through a joint processing module for the final diagnosis and severity prediction. This module accumulates the volumetric features from the generalized feature representation of each slice as well as introduces an effective fusion of features to generate the corresponding representative feature vector of the CT-volume. Moreover, this module plays an influential role in the proper selection of features especially in the early stage of infection when few of the slices contain infected lesions. To facilitate the feature selection process, the processing of severity based features and diagnostic features are isolated. In Fig.~\ref{f1}, separate volumetric feature aggregation and fusion modules are integrated to separately optimize the diagnostic and severity features.
Though similar operational modules are employed in both of these cases, another stage of attention-gating operations is employed to guide the effective slice-wise features in these operational modules with different objectives (shown in Fig.~\ref{f1}).
This module is schematically presented in Fig.~\ref{f6}. Firstly, the volumetric feature accumulation is carried out to produce the aggregated feature vector $F_{agg}$ from the regional features ($F_{reg}$) of all slices.
Thereupon, the fusion scheme is employed utilizing dilated convolutions~\cite{dil} which provides the opportunity to explore features from diverse receptive areas. Firstly, a pointwise convolution $(1\times 1)$ is carried out for depth reduction of the aggregated vector $F_{agg}$. Subsequently, several dilated convolutions are operated with varying dilation rates for the effective fusion of features, and outputs of these convolutions are processed through another stage of aggregation, convolutional filtering, and global pooling operations to generate a 1D-representational feature vector. Finally, several fully connected layer operations are employed for generating the final prediction for a specific CT-volume.

\subsection{Loss Functions}
The optimization of the whole process is divided into two phases where the TA-SegNet is optimized in the first phase and joint optimization of the diagnostic and severity prediction tasks are carried out in the second phase utilizing the optimized TA-SegNet from phase-1.  A focal Tversky loss function ($\mathscr{L}_{FTL}$) is proposed in~\cite{ftl} utilizing the Tversky index that performs well over a large range of applications which is used as the objective function to optimize TA-SegNet.

In general, both the COVID diagnosis and severity predictions are defined as binary-classification tasks, where normal/disease classes are considered for diagnosis while mild/severe classes are considered for severity predictions. For joint optimization of the diagnosis and severity prediction, an objective loss function ($L_{obj}$) is defined by combining the objective loss functions for diagnosis ($L_d$) and severity prediction ($L_s$). The severity prediction task will only be initiated for the infected volumes where $y_d=1$, while for the normal cases ($y_d=0$), this task is ignored. However, the diagnosis task is carried out for all normal/infectious volumes. Hence, the objective loss function ($L_{obj}$) can be expressed as:
\begin{align}
    L_{obj} &= L_d(\mathbf{Y_d}, \mathbf{Y_d^p}) + L_s(\mathbf{Y_d}, \mathbf{Y_s}, \mathbf{Y_s^p}) \nonumber \\
    &= \frac{1}{M} \sum_{i=1}^M  \mathscr{L}_B(y_{d,i}, y_{d,i}^p) + \frac{1}{M_I} \sum_{i=1}^{M_I} y_{d,i} \mathscr{L}_B(y_{s,i}, y_{s,i}^p)
\end{align}
where $\mathbf{Y_d}$ and $\mathbf{Y_s}$ represent the set of diagnosis and severity ground truths while $\mathbf{Y^p_{d}, Y^p_{s}}$ represent the corresponding set of predictions, $\mathscr{L}_B$ denotes binary cross-entropy loss, $M$ denotes the total number of CT-volumes, and $M_I$ represents the total number of infected volumes. Hence, the proposed CovTANet network can be effectively optimized for joint segmentation, diagnosis, and severity predictions of COVID-19 utilizing this two phase optimization scheme.

\begin{figure*}[t]
    \centering
    \includegraphics[width=\linewidth]{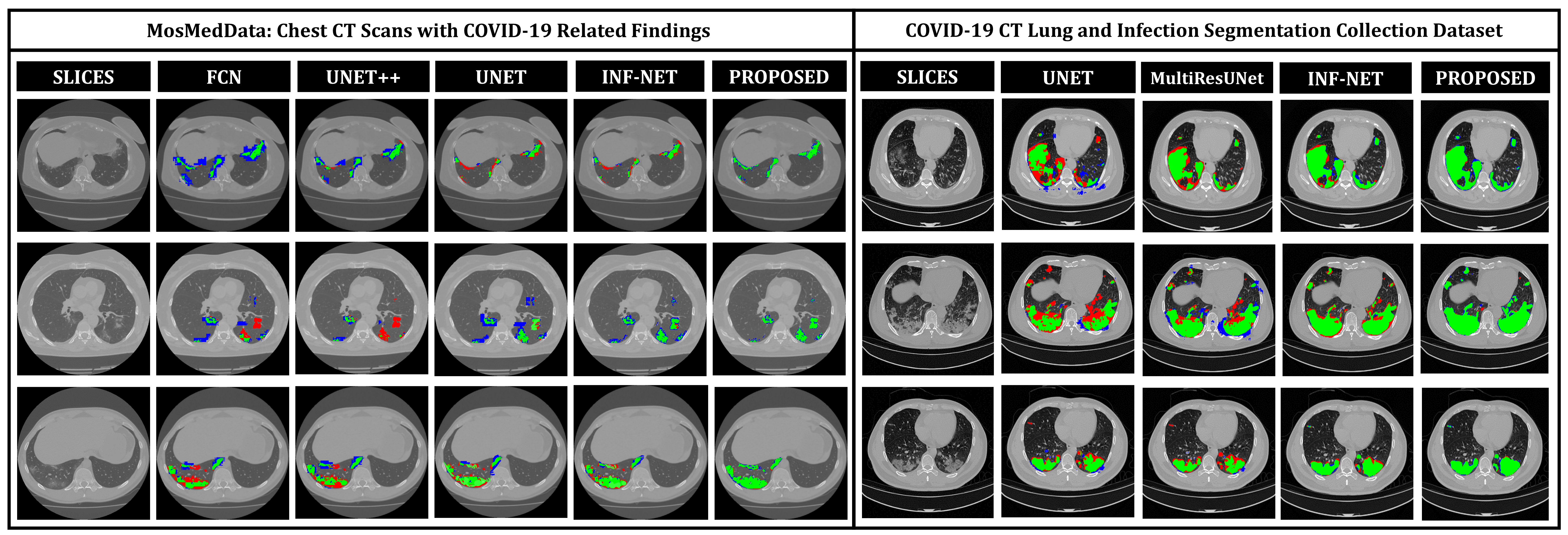}
    \caption{\textbf{Visualization of the lesion segmentation performance of some of the state-of-the-art networks in MosMedData~\cite{d1} and  Dataset-2~\cite{d2}. Here, \textcolor{green}{`green'} denotes the true positive (TP) region, \textcolor{blue}{`blue'} denotes the false positive region, and \textcolor{red}{`red'} denotes the false negative regions. }}
    \label{f7}
\end{figure*}

\section{Results and Discussions}
In this section, results obtained from extensive experimentation on a publicly available dataset are presented and discussed from diverse perspectives to validate the effectiveness of the proposed scheme.

\begin{table}[t]
\centering
\caption{{Performance (Mean $\pm$ Standard Deviation) of the Ablation Study of the Proposed TA-SegNet on MosMedData}}
\label{t1}
\scalebox{0.75}{
\begin{tabular}{|c|c|c|c|c|c|c|}
\hline
\multicolumn{1}{|l|}{\textbf{Version}} & \textbf{\begin{tabular}[c]{@{}c@{}}EfficientNet\\ Backbone\end{tabular}} & \textbf{\begin{tabular}[c]{@{}c@{}}Encoder\\ TAU Unit\end{tabular}} & \textbf{\begin{tabular}[c]{@{}c@{}}Decoder\\ TAU Unit\end{tabular}} & \textbf{\begin{tabular}[c]{@{}c@{}}Encoder\\ in Fusion\end{tabular}} & \textbf{\begin{tabular}[c]{@{}c@{}}Decoder\\ in Fusion\end{tabular}} & \textbf{Dice(\%)}              \\ \hline
V1                                     & \xmark                                                                   & \xmark                                                               & \xmark                                                               & \xmark                                                                & \xmark                                                                & 50.5$\pm$0.26                      \\ \hline
V2                                     & \xmark                                                                    & \xmark                                                               & \xmark                                                               & \xmark                                                                & \cmark                                                                & 52.4$\pm$0.17                      \\ \hline
V3                                     & \xmark                                                                    & \cmark                                                               & \xmark                                                               & \xmark                                                                & \xmark                                                                & 54.9$\pm$0.14                      \\ \hline

V4                                     & \xmark                                                                    & \cmark                                                               & \xmark                                                               & \xmark                                                                & \xmark                                                                & 54.6$\pm$0.14                      \\ \hline
V5                                     & \xmark                                                                    & \xmark                                                               & \cmark                                                               & \xmark                                                                & \xmark                                                                & 53.4$\pm$0.19                      \\ \hline
V6                                     & \xmark                                                                    & \cmark                                                               & \cmark                                                               & \xmark                                                                & \xmark                                                                & 57.1$\pm$0.33                      \\ \hline
V7                                     & \xmark                                                                    & \cmark                                                               & \cmark                                                               & \cmark                                                                & \cmark                                                                & 60.2$\pm$0.26                      \\ \hline
V8                                     & \cmark                                                                    & \cmark                                                               & \cmark                                                               & \cmark                                                                & \cmark                                                                & 62.3$\pm$0.18                      \\ \hline
\end{tabular}}
\end{table}

\begin{table}[t]
\centering
\caption{Comparison of Performances with Other the State-of-the-Art Networks on COVID Lesion Segmentation on MosMedData}
\label{t2}
\scalebox{0.85}{
\begin{tabular}{|c|c|c|c|c|}
\hline
\textbf{Networks}       & \textbf{Sensitivity(\%)} & \textbf{Precision(\%)} & \textbf{Dice(\%)}  & \textbf{IoU(\%)}   \\ \hline
FCN~\cite{fcn}                     & 78.8$\pm$0.23             & 58.9$\pm$0.16              & 35.6$\pm$0.36          & 29.3$\pm$0.45          \\ \hline
Unet~\cite{unet}                    & 94.3$\pm$0.34             & 74.4$\pm$0.32              & 50.5$\pm$0.26          & 40.3$\pm$0.23          \\ \hline
Vnet~\cite{vnet}                    & 84.5$\pm$0.42             & 64.6$\pm$0.54              & 40.2$\pm$0.33          & 36.4$\pm$0.26          \\ \hline
Unet++~\cite{unet++}                  & 78.1$\pm$0.15             & 65.1$\pm$0.25              & 37.2$\pm$0.27          & 33.3$\pm$0.32          \\ \hline
CPF-Net~\cite{cpfnet}                 & 82.4$\pm$0.25             & 71.3$\pm$0.29              & 48.9$\pm$0.21          & 37.6$\pm$0.38          \\ \hline
COPLE-Net~\cite{cop}     & 85.5$\pm$0.18             & 73.1$\pm$0.20              & 51.1$\pm$0.21          & 41.2$\pm$0.38          \\ \hline
Mini-SegNet~\cite{mini} & 81.5$\pm$0.25             & 69.1$\pm$0.19              & 43.7$\pm$0.23          & 35.2$\pm$0.38          \\ \hline
Inf-Net~\cite{inf}                 & 92.8$\pm$0.27             & 76.9$\pm$0.34              & 51.8$\pm$0.31          & 41.6$\pm$0.27          \\ \hline
\textbf{TA-SegNet (Prop.)} & \textbf{99.6$\pm$0.09}    & \textbf{84.8$\pm$0.26}     & \textbf{62.3$\pm$0.18} & \textbf{51.7$\pm$0.29} \\ \hline
\end{tabular}
}
\end{table}

\subsection{Dataset Description}
This study is conducted using ``MosMedData: Chest CT Scans with COVID-19 Related Findings''~\cite{d1}, one of the largest publicly available datasets in this domain. The dataset, being collected from the hospitals in Moscow, Russia, contains 1110 anonymized CT-volumes with severity annotated COVID-19 related findings, as well as without such findings. Each one of the 1110 CT-volumes is acquired from different persons and 30-46 slices per patient are available. Pixel annotations of the COVID lesions are provided for 50 CT-volumes which are used for training and evaluation of the proposed TA-SegNet. For carrying out the diagnosis and severity prediction tasks, all the CT-volumes are divided into normal, mild ($<$25\% lung parenchyma) and severe ($>$25\% lung parenchyma) infection categories. 

In addition, to validate the lesion segmentation performance of our method, ``COVID-19 CT Lung and Infection Segmentation Collection dataset''~\cite{d2} is employed as a secondary dataset that contains 20 CT volumes (Average slices 176 per volume) collected from 19 different patients with pixel-annotated lung and infection regions labeled by two expert radiologists. However, this dataset does not contain severity-based annotation and therefore, is mainly incorporated for additional experimentation on the segmentation performance using TA-SegNet.

%This data set was collected from the public data set “MosMedData: Chest CT Scans with COVID-19 Related Findings”,  which is, at the time of writing this paper, the only publicly available 3D volume set with 2-C and 2-A annotations, that can be used for segmentation and severity detection. The used dataset consists of 1110 volumes of computer tomography(CT) scans of Human Lungs with COVID-19 related findings, as well as without such findings. A subset of 50 volumes was annotated with binary pixel masks, depicting ground-glass opacifications and consolidations. These CT scan volumes were obtained between 1st of March 2020 and 25th of April 2020, from medical hospitals in Moscow, Russia. 

%The scans were taken using Canon(Toshiba) Aquillion 64 unit, following standard scanner protocols. While taking the scans, the inter-slice distance was 0.8mm, but the public data set contains every 10th image, making the inter-slice distance 8mm. For severity classification, the data set follows a visual semi-quantitative scale adopted in the Russian Federation and Moscow [Morozov et al. (2020d)]. The patients are categorized as CT 0 (healthy patients with no COVID symptoms), CT 1 (COVID-19 symptoms with 0-25\% lung parenchyma), CT 2 (COVID-19 symptoms with 25-50\% lung parenchyma), CT 3 (COVID-19 symptoms with 50-75\% lung parenchyma), CT 4 (COVID-19 symptoms with above 75\% lung parenchyma). 
\subsection{Experimental Setup}
%Intel\textregistered\  Xeon\textregistered \ $D-1653N$ CPU @$2.80$GHz with $12$M Cache and $8$ cores along with $24$ GB RAM is used for experimentation. $1\times$ NVIDIA RTX $2080$ Ti GPU running at $1770$ MHz with $24$ GB GDDR6 memory having $4608$ CUDA cores is deployed for hardware acceleartion.
With a five-fold cross-validation scheme over the MosMed dataset, all the experimentations have been implemented on the google cloud platform with NVIDIA P-100 GPU as the hardware accelerator. For evaluation of the segmentation performance, some of the traditional metrics are used, such as accuracy, precision, dice score, and intersection-over-union (IoU) score, while for assessing the severity classification performance, accuracy, sensitivity, specificity, and F1-score are used. The Adam optimizer is employed with an initial learning rate of $10^{-5}$ which is decayed at a rate of 0.99 after every 10 epochs.

\begin{table}[b]
\centering
\caption{Comparison of Performances with Other the State-of-the-Art Networks on COVID Lesion Segmentation on Dataset-2}
\label{t4}
\scalebox{0.83}{
\begin{tabular}{|c|c|c|c|c|}
\hline
\textbf{Network}        & \textbf{Sensitivity(\%)} & \textbf{Specificity(\%)} & \textbf{Dice(\%)} & \textbf{IoU(\%)} \\ \hline
\textbf{Unet~\cite{unet}}           & 75.9$\pm$0.34            & 88.9$\pm$0.12            & 79.3$\pm$0.26     & 74.9$\pm$0.18    \\ \hline

\textbf{MultiResUnet~\cite{mru}}   & 77.2$\pm$0.33            & 90.3$\pm$0.24            & 82.7$\pm$0.28     & 77.4$\pm$0.15    \\ \hline
\textbf{Attention-Unet~\cite{att}} & 81.1$\pm$0.29            & 92.2$\pm$0.11            & 85.1$\pm$0.14     & 79.6$\pm$0.28    \\ \hline
\textbf{CPF-Net~\cite{cpfnet}}        & 78.9$\pm$0.27            & 91.7$\pm$0.14            & 84.4$\pm$0.25     & 79.3$\pm$0.25    \\ \hline
\textbf{Gated-Unet~\cite{gated}}     & 81.4$\pm$0.24            & 92.5$\pm$0.19            & 85.6$\pm$0.19     & 80.2$\pm$0.16    \\ \hline
\textbf{Inf-Net~\cite{inf}}        & 82.7$\pm$0.26            & 94.8$\pm$0.21            & 86.9$\pm$0.34     & 81.1$\pm$0.18    \\ \hline
\textbf{TA-SegNet (ours)} & 88.5$\pm$0.22            & 98.9$\pm$0.14            & 90.2$\pm$0.17     & 86.4$\pm$0.19    \\ \hline
\end{tabular}}
\end{table}

% \usepackage{multirow}
\begin{table*}[t]
\centering
\caption{{Comparison of Performances in the Joint Diagnosis and Severity Prediction of COVID-19 with Different Networks on MosMedData}}
\label{t3}
\scalebox{0.75}{
\begin{tabular}{|c|c|c|c|c|c|c|c|c|c|c|c|c|}
\hline
\multirow{3}{*}{\textbf{Network}} & \multicolumn{8}{c|}{\textbf{Diagnostic Prediction}}                                                                                                                   & \multicolumn{4}{c|}{\textbf{Severity Prediction}}                                 \\ \cline{2-13} 
                                  & \multicolumn{4}{c|}{\textbf{Normal vs. Mild}}                                     & \multicolumn{4}{c|}{\textbf{Normal vs. Severe}}                                   & \multicolumn{4}{c|}{\textbf{Mild vs. Severe}}                                     \\ \cline{2-13} 
                                  & \textbf{Sen.(\%)}  & \textbf{Spec.(\%)} & \textbf{Acc.(\%)}  & \textbf{F1(\%)}    & \textbf{Sen.(\%)}  & \textbf{Spec.(\%)} & \textbf{Acc.(\%)}  & \textbf{F1(\%)}    & \textbf{Sen.(\%)}  & \textbf{Spec.(\%)} & \textbf{Acc.(\%)}  & \textbf{F1(\%)}    \\ \hline
VGG-19                            & 54.4$\pm$0.37          & 63.4$\pm$0.28          & 58.4$\pm$0.32          & 58.6$\pm$0.33          & 63.4$\pm$0.32          & 70.8$\pm$0.37          & 65.9$\pm$0.28          & 66.9$\pm$0.39          & 62.7$\pm$0.25          & 65.5$\pm$0.23          & 61.9$\pm$0.25          & 64.1$\pm$0.28          \\ \hline
ResNet-50                         & 61.1$\pm$0.29          & 65.7$\pm$0.33          & 62.5$\pm$0.44          & 63.3$\pm$0.37          & 66.5$\pm$0.26          & 69.3$\pm$0.31          & 69.1$\pm$0.17          & 67.8$\pm$0.33          & 61.1$\pm$0.36          & 63.8$\pm$0.29          & 64.8$\pm$0.34          & 62.4$\pm$0.37          \\ \hline
Xception                          & 56.8$\pm$0.38          & 57.9$\pm$0.41          & 59.9$\pm$0.27          & 57.3$\pm$0.43          & 64.8$\pm$0.19          & 67.2$\pm$0.26          & 66.7$\pm$0.21          & 65.9$\pm$0.25          & 62.9$\pm$0.33          & 65.1$\pm$0.32          & 63.1$\pm$0.22          & 63.9$\pm$0.34          \\ \hline
DenseNet121                       & 59.7$\pm$0.27          & 64.6$\pm$0.21          & 61.1$\pm$0.39          & 62.1$\pm$0.28          & 65.1$\pm$0.23          & 70.1$\pm$0.19          & 67.8$\pm$0.29          & 67.5$\pm$0.25          & 60.2$\pm$0.28          & 64.4$\pm$0.21          & 60.6$\pm$0.26          & 62.2$\pm$0.27           \\ \hline
InceptionV3                       & 60.4$\pm$0.31          & 62.1$\pm$0.38          & 59.3$\pm$0.23          & 61.2$\pm$0.35          & 66.6$\pm$0.28          & 69.8$\pm$0.24          & 66.2$\pm$0.32          & 68.2$\pm$0.31          & 62.8$\pm$0.29          & 67.9$\pm$0.35          & 61.4$\pm$0.19          & 65.3$\pm$0.33          \\ \hline
\textbf{CovTANet (ours)}                 & \textbf{83.8$\pm$0.25} & \textbf{90.3$\pm$0.27} & \textbf{85.2$\pm$0.19} & \textbf{86.9$\pm$0.22} & \textbf{93.9$\pm$0.13} & \textbf{96.6$\pm$0.11} & \textbf{95.8$\pm$0.17} & \textbf{94.2$\pm$0.21} & \textbf{90.9$\pm$0.16} & \textbf{93.4$\pm$0.23} & \textbf{91.7$\pm$0.19} & \textbf{92.1$\pm$0.11} \\ \hline
\end{tabular}
}
\end{table*}

% \begin{table}[t]
% \centering
% \caption{Comparison of Performances with Other the State-of-the-Art Networks on COVID Lesion Segmentation on Dataset-2}
% \label{t4}
% \scalebox{0.83}{
% \begin{tabular}{|c|c|c|c|c|}
% \hline
% \textbf{Network}        & \textbf{Sensitivity(\%)} & \textbf{Specificity(\%)} & \textbf{Dice(\%)} & \textbf{IoU(\%)} \\ \hline
% \textbf{Unet~\cite{unet}}           & 75.9$\pm$0.34            & 88.9$\pm$0.12            & 79.3$\pm$0.26     & 74.9$\pm$0.18    \\ \hline

% \textbf{MultiResUnet~\cite{mru}}   & 77.2$\pm$0.33            & 90.3$\pm$0.24            & 82.7$\pm$0.28     & 77.4$\pm$0.15    \\ \hline
% \textbf{Attention-Unet~\cite{att}} & 81.1$\pm$0.29            & 92.2$\pm$0.11            & 85.1$\pm$0.14     & 79.6$\pm$0.28    \\ \hline
% \textbf{CPF-Net~\cite{cpfnet}}        & 78.9$\pm$0.27            & 91.7$\pm$0.14            & 84.4$\pm$0.25     & 79.3$\pm$0.25    \\ \hline
% \textbf{Gated-Unet~\cite{gated}}     & 81.4$\pm$0.24            & 92.5$\pm$0.19            & 85.6$\pm$0.19     & 80.2$\pm$0.16    \\ \hline
% \textbf{Inf-Net~\cite{inf}}        & 82.7$\pm$0.26            & 94.8$\pm$0.21            & 86.9$\pm$0.34     & 81.1$\pm$0.18    \\ \hline
% \textbf{TA-SegNet (ours)} & 88.5$\pm$0.22            & 98.9$\pm$0.14            & 90.2$\pm$0.17     & 86.4$\pm$0.19    \\ \hline
% \end{tabular}}
% \end{table}

\subsection{Analysis of the Segmentation Performance} 
Firstly, ablation studies are carried out to validate the effectiveness of different modules in TA-SegNet. Afterwards, the performance of the best performing variant is compared with other networks from qualitative and quantitative perspectives. 
\subsubsection{Ablation study}
Traditional Unet network has been used as a baseline model (V1) and five other schemes/modules have been incorporated in the baseline model to analyze the contribution of different modules in the performance improvement of the proposed TA-SegNet (V8). For ease of comparison, only Dice score is used as it is the most widely used metric for segmentation tasks.
%For carrying out ablation study, these units are removed from the TA-Unet structure.  
From Table~\ref{t1}, it can be noted that the encoder TAUs (V4) provide 4.1\% improvement of the Dice score from the baseline, while the decoder TAUs (V5) provide a 2.9\% improvement and when both these are combined (V6), 6.6\% improvement is achieved. As the encoder TAUs contribute significantly to the reduction of semantic gaps with the corresponding decoder feature maps, while the decoder TAU units guide the decoded feature maps with finer details for better generalization of multi-scale features, considerable performance improvement is achieved when employed in combination. Moreover, all the multi-scale feature maps generated from various encoder levels are guided to the reconstruction process through a deep fusion scheme along with the multi-scale decoded feature maps. The integration of these multi-scale features from the encoder-decoder modules in the fusion process (V3) contributes to the efficient reconstruction, and  4.4\% improvement of Dice score is achieved over the baseline. Moreover, 9.7\% improvement of Dice score is achieved when the fusion scheme is combined with two-stage TAU-units (V7). Additionally, for introducing transfer learning, pre-trained models on the ImageNet database can be used as the backbone of the encoder module of the TA-SegNet similar to most other segmentation networks. It should be noted that with the pre-trained EfficientNet network as the backbone of the encoder module (V8), the performance gets improved by 2.1\% compared to the TA-SegNet framework without such backbone (V7).

\subsubsection{Quantitative analysis}
In Table~\ref{t2}, performances of some of the state-of-the-art networks are summarized. It should be noticed that the proposed TA-SegNet outperforms all the methods compared by a considerable margin in all the metrics. Using the proposed framework, 11.8\% improvement of Dice score over Unet, and 26.7\% improvement of Dice score over the FCN have been achieved. Furthermore, our network improves the dice score of the second-best method (Inf-Net) by about 10.5\%,  which intuitively indicates its excellent capabilities over the rest of the models. The robustness of the proposed scheme and the enhanced capability of our model in terms of infected region identification is further demonstrated by the high sensitivity score (99.6\%) reported. This signifies the fact that the model integrates the symmetric encoding-decoding strategy of Unet as well as exploits the parallel optimization advantages of FCN that provides this large improvement. Most other state-of-the-art variants of the Unet provide sub-optimal performances for increasing complexity considerably that makes the optimization difficult in most of the challenging cases. However, due to the smaller amount of infections in the annotated CT-volumes used for training and optimization of the segmentation networks, a higher amount of false positives have been generated in most of the networks which reduced the precision. The proposed TA-SegNet has considerably reduced the false positives along with false negatives and has improved both precision and sensitivity.

\subsubsection{Qualitative analysis} 
In Fig.~\ref{f7}, qualitative representations of the segmentation performances of different networks are shown in some challenging conditions. The comparable dimensions of the small infected regions and the arteries, veins embedded in the thorax cavity with varying anatomical appearance might be attributed to the large occurrences of the false positives. It is evident that most other networks struggle to extract the complicated, scattered, and diffused COVID-19 lesions, while the proposed TA-SegNet considerably improves the segmentation performance in such challenging conditions. This depiction conforms to the fact that our network can correctly segment both of the large and small infected regions. Furthermore, our framework consistently demonstrates almost non-existent false negatives compared to the other models while considerably reducing the false positive predictions as it can distinguish sharper details of the lesions and effectively perform for early diagnosis of the infection.

%for lesion for lesion segmentation, early diagnosis, and severity predictions of COVID-19

\subsection{Performance on Secondary Dataset}
In Table~\ref{t4}, quantitative performances on Dataset-2~\cite{d2} for different networks are summarized. It should be noticed that the proposed TA-SegNet provides the best achievable performances with $90.2\%$ mean Dice score while providing $10.9\%$ improvement over Unet. In Fig.~\ref{f7}, the qualitative performance analyses are provided on several challenging examples that further demonstrate the effectiveness of the proposed TA-SegNet over other traditional networks with large reductions in both false positives and false negatives. However, it should be mentioned that this dataset contains mostly higher levels of infections in the CT volumes on average that makes the learning and optimization more favorable compared to the MosMedData, and thus, comparatively higher segmentation performances have been achieved.

\subsection{Analysis of the Joint Classification Performance}
In Table~\ref{t3}, the performances obtained from the joint diagnosis and severity prediction tasks are summarized. To analyze the effectiveness of the proposed multi-phase optimization scheme, some of the state-of-the-art networks are also evaluated for the slice-wise processing of the CT-volumes in the joint-classification scheme discarding the TA-SegNet. 

\subsubsection{Diagnostic prediction performance analysis}
The diagnosis performances with mild and severe cases of COVID-19 are separately reported to distinguish the early diagnosis performance. The proposed CovTANet provides 85.2\% accuracy in isolating the COVID patients even with mild symptoms, while the accuracy is as high as 95.8\% when the CT volumes contain severe infections. However, the other networks operating without the TA-SegNet noticeably suffer especially in the early diagnosis phase, as it is difficult to isolate the small infection patches from the CT-volume.
Hence, it can be interpreted that this high early diagnostic accuracy of CovTANet is significantly contributed by the multi-phase optimization scheme that incorporates the highly optimized TA-SegNet for extracting the most effective lesion features to mitigate the effect of redundant healthy parts. 

% \begin{figure*}[t]
%     \centering
%     \includegraphics[scale=0.7]{vis2.png}
%     \caption{\textbf{Visualization of the lesion segmentation performance of some of the state-of-the-art networks in Dataset-2~\cite{d2}. Here, \textcolor{green}{`green'} denotes the true positive (TP) region, \textcolor{blue}{`blue'} denotes the false positive region, and \textcolor{red}{`red'} denotes the false negative regions. }}
%     \label{f9}
% \end{figure*}

\subsubsection{Severity prediction performance analysis}
In the joint optimization process based on the amount of infected lung parenchymas, mild and severe patients are also categorized. Despite the additional challenges regarding the isolation and quantification of the abnormal tissues, the proposed scheme generalizes the problem quite well which provides 91.7\% accuracy in categorizing mild and severe patients. It should be noted that the highest achievable severity prediction accuracy with a traditional network is 64.8\% (using ResNet50) with considerably smaller results in most other metrics. Traditional network directly operates on the whole CT-volume to extract effective features for severity prediction which makes the task more complicated. Whereas, the proposed hybrid CovTANet with multiphase optimization effectively integrates features regarding infections from the TA-SegNet for considerably simplifying the feature extraction process in the joint-classification process that results in higher accuracy.

\section{Conclusion and Future Works}
In this study, a multi-phase optimization scheme is proposed with a hybrid neural network (CovTANet) 
where an efficient lesion segmentation network is integrated into a complete optimization framework for joint diagnosis and severity prediction of COVID-19 from CT-volume. 
The tri-level attention mechanism and parallel optimization of multi-scale encoded-decoded feature maps which are introduced in the segmentation network (TA-SegNet) have improved the lesion segmentation performance substantially. Moreover, the effective integration of features from the optimized TA-SegNet is found to be extremely beneficial in diagnosis and severity prediction by de-emphasizing the effects of redundant features from the whole CT-volumes. It is also shown that the proposed joint classification scheme not only provides better diagnosis at severe infection stages but also is capable of early diagnosis of patients having mild infections with outstanding precision. Furthermore, considerable performances have been achieved in severity screening that would facilitate a faster clinical response to substantially reduce the probable damages.  
Nonetheless, a further study should be carried out considering patients from diverse geographic locations to understand the mutation and evolution of this deadly virus where the proposed hybrid network is supposed to be very effective.
The proposed scheme can be a valuable tool for the clinicians to combat this pernicious disease through faster-automated mass-screening.  % with significantly higher precision.

\bibliographystyle{IEEEtran}
\bibliography{ref}

\begin{thebibliography}{10}
\providecommand{\url}[1]{#1}
\csname url@samestyle\endcsname
\providecommand{\newblock}{\relax}
\providecommand{\bibinfo}[2]{#2}
\providecommand{\BIBentrySTDinterwordspacing}{\spaceskip=0pt\relax}
\providecommand{\BIBentryALTinterwordstretchfactor}{4}
\providecommand{\BIBentryALTinterwordspacing}{\spaceskip=\fontdimen2\font plus
\BIBentryALTinterwordstretchfactor\fontdimen3\font minus
  \fontdimen4\font\relax}
\providecommand{\BIBforeignlanguage}[2]{{%
\expandafter\ifx\csname l@#1\endcsname\relax
\typeout{** WARNING: IEEEtran.bst: No hyphenation pattern has been}%
\typeout{** loaded for the language `#1'. Using the pattern for}%
\typeout{** the default language instead.}%
\else
\language=\csname l@#1\endcsname
\fi
#2}}
\providecommand{\BIBdecl}{\relax}
\BIBdecl

\bibitem{b1}
G.~Meyerowitz-Katz and L.~Merone, ``A systematic review and meta-analysis of
  published research data on {COVID-19} infection fatality rates,''
  \emph{International Journal of Infectious Diseases}, vol. 101, pp. 138 --
  148, 2020.

\bibitem{rt}
Y.~Fang, H.~Zhang, J.~Xie, M.~Lin, L.~Ying, P.~Pang, and W.~Ji, ``Sensitivity
  of chest {CT} for {COVID-19}: comparison to {RT-PCR},'' \emph{Radiology},
  vol. 296, no.~2, pp. E115--E117, 2020.

\bibitem{ai}
F.~Shi, J.~Wang, J.~Shi, Z.~Wu, Q.~Wang, Z.~Tang, K.~He, Y.~Shi, and D.~Shen,
  ``Review of artificial intelligence techniques in imaging data acquisition,
  segmentation and diagnosis for {COVID-19},'' \emph{IEEE Reviews in Biomedical
  Engineering}, 2020.

\bibitem{abdel3}
M.~Abdel-Basset, V.~Chang, and N.~A. Nabeeh, ``An intelligent framework using
  disruptive technologies for {COVID-19} analysis,'' \emph{Technological
  Forecasting and Social Change}, p. 120431, 2020.

\bibitem{ct1}
L.~Li, L.~Qin, Z.~Xu, Y.~Yin, X.~Wang, B.~Kong, J.~Bai, Y.~Lu, Z.~Fang, Q.~Song
  \emph{et~al.}, ``Artificial intelligence distinguishes {COVID-19} from
  community acquired pneumonia on chest {CT},'' \emph{Radiology}, vol. 296,
  no.~2, pp. E65--E71, 2020.

\bibitem{ct2}
L.~Huang, R.~Han, T.~Ai, P.~Yu, H.~Kang, Q.~Tao, and L.~Xia, ``Serial
  quantitative chest {CT} assessment of {COVID-19}: Deep-learning approach,''
  \emph{Radiology: Cardiothoracic Imaging}, vol.~2, no.~2, p. e200075, 2020.

\bibitem{abdel2}
M.~Abdel-Basset, V.~Chang, H.~Hawash, R.~K. Chakrabortty, and M.~Ryan,
  ``{FSS-2019-nCov}: A deep learning architecture for semi-supervised few-shot
  segmentation of {COVID-19} infection,'' \emph{Knowledge-Based Systems}, p.
  106647, 2020.

\bibitem{x1}
T.~Mahmud, M.~A. Rahman, and S.~A. Fattah, ``{CovXNet}: A multi-dilation
  convolutional neural network for automatic {COVID-19} and other pneumonia
  detection from chest {X-ray} images with transferable multi-receptive feature
  optimization,'' \emph{Computers in Biology and Medicine}, p. 103869, 2020.

\bibitem{abdel}
M.~Abdel-Basset, V.~Chang, and R.~Mohamed, ``{HSMA\_WOA: A hybrid novel slime
  mould algorithm with whale optimization algorithm for tackling the image
  segmentation problem of chest {X-ray} images},'' \emph{Applied Soft
  Computing}, vol.~95, p. 106642, 2020.

\bibitem{early}
J.~P. Kanne, B.~P. Little, J.~H. Chung, B.~M. Elicker, and L.~H. Ketai,
  ``Essentials for radiologists on {COVID-19}: an update—radiology scientific
  expert panel,'' \emph{Radiology}, vol. 296, no.~2, pp. E113--E114, 2020.

\bibitem{severity}
J.~T. Wu, K.~Leung, M.~Bushman, N.~Kishore, R.~Niehus, P.~M. de~Salazar, B.~J.
  Cowling, M.~Lipsitch, and G.~M. Leung, ``Estimating clinical severity of
  {COVID-19} from the transmission dynamics in {Wuhan, China},'' \emph{Nature
  Medicine}, vol.~26, no.~4, pp. 506--510, 2020.

\bibitem{inf}
D.~P. {Fan}, T.~{Zhou}, G.~P. {Ji}, Y.~{Zhou}, G.~{Chen}, H.~{Fu}, J.~{Shen},
  and L.~{Shao}, ``{Inf-Net}: Automatic {COVID-19} lung infection segmentation
  from {CT} images,'' \emph{IEEE Transactions on Medical Imaging}, vol.~39,
  no.~8, pp. 2626--2637, 2020.

\bibitem{mini}
Y.~Qiu, Y.~Liu, and J.~Xu, ``Miniseg: An extremely minimum network for
  efficient {COVID-19} segmentation,'' \emph{arXiv preprint arXiv:2004.09750},
  2020.

\bibitem{cop}
G.~Wang, X.~Liu, C.~Li, Z.~Xu, J.~Ruan, H.~Zhu, T.~Meng, K.~Li, N.~Huang, and
  S.~Zhang, ``A noise-robust framework for automatic segmentation of {COVID-19}
  pneumonia lesions from {CT} images,'' \emph{IEEE Transactions on Medical
  Imaging}, vol.~39, no.~8, pp. 2653--2663, 2020.

\bibitem{new2}
L.~Huang, R.~Han, T.~Ai, P.~Yu, H.~Kang, Q.~Tao, and L.~Xia, ``Serial
  quantitative chest {CT} assessment of {COVID-19}: Deep-learning approach,''
  \emph{Radiology: Cardiothoracic Imaging}, vol.~2, no.~2, p. e200075, 2020.

\bibitem{unet}
O.~Ronneberger, P.~Fischer, and T.~Brox, ``U-net: Convolutional networks for
  biomedical image segmentation,'' in \emph{2015 18th International Conference
  on Medical image computing and computer-assisted intervention}, 2015, pp.
  234--241.

\bibitem{unet++}
Z.~Zhou, M.~M.~R. Siddiquee, N.~Tajbakhsh, and J.~Liang, ``Unet++: Redesigning
  skip connections to exploit multiscale features in image segmentation,''
  \emph{IEEE Transactions on Medical Imaging}, vol.~39, no.~6, pp. 1856--1867,
  2019.

\bibitem{new1}
A.~Bernheim, X.~Mei, M.~Huang, Y.~Yang, Z.~A. Fayad, N.~Zhang, K.~Diao, B.~Lin,
  X.~Zhu, K.~Li \emph{et~al.}, ``Chest {CT} findings in coronavirus disease-19
  {(COVID-19)}: relationship to duration of infection,'' \emph{Radiology}, vol.
  295, no.~3, p. 200463, 2020.

\bibitem{attention}
A.~Sinha and J.~Dolz, ``Multi-scale self-guided attention for medical image
  segmentation,'' \emph{IEEE Journal of Biomedical and Health Informatics},
  2020.

\bibitem{attention2}
A.~Vaswani, N.~Shazeer, N.~Parmar, J.~Uszkoreit, L.~Jones, A.~N. Gomez,
  {\L}.~Kaiser, and I.~Polosukhin, ``Attention is all you need,'' in
  \emph{Advances in Neural Information Processing Systems}, 2017, pp.
  5998--6008.

\bibitem{squeeze}
J.~Hu, L.~Shen, and G.~Sun, ``Squeeze-and-excitation networks,'' in \emph{2018
  IEEE conference on computer vision and pattern recognition}, 2018, pp.
  7132--7141.

\bibitem{attention3}
Z.~Tan, Y.~Yang, J.~Wan, H.~Hang, G.~Guo, and S.~Z. Li, ``Attention-based
  pedestrian attribute analysis,'' \emph{IEEE Transactions on Image
  Processing}, vol.~28, no.~12, pp. 6126--6140, 2019.

\bibitem{dil}
F.~Yu and V.~Koltun, ``Multi-scale context aggregation by dilated
  convolutions,'' \emph{arXiv preprint arXiv:1511.07122}, 2015.

\bibitem{ftl}
N.~Abraham and N.~M. Khan, ``A novel focal tversky loss function with improved
  attention {U-net} for lesion segmentation,'' in \emph{2019 IEEE 16th
  International Symposium on Biomedical Imaging}, 2019, pp. 683--687.

\bibitem{d1}
``{{MosMedData}: Chest {CT} Scans with {COVID-19} Related Findings},'' 2020,
  accessed: 28 April, 2020. [online]. Available:
  \url{https://mosmed.ai/datasets/covid19_1110}.

\bibitem{d2}
``{{COVID-19 CT} Lung and Infection segmentation dataset},'' 2020, accessed: 16
  October, 2020. [online]. Available: \url{https://zenodo.org/record/3757476}.

\bibitem{fcn}
J.~Long, E.~Shelhamer, and T.~Darrell, ``Fully convolutional networks for
  semantic segmentation,'' in \emph{2015 IEEE 28th conference on computer
  vision and pattern recognition}, 2015, pp. 3431--3440.

\bibitem{vnet}
F.~Milletari, N.~Navab, and S.-A. Ahmadi, ``V-net: Fully convolutional neural
  networks for volumetric medical image segmentation,'' in \emph{{2016 4th
  International Conference on 3D Vision}}, 2016, pp. 565--571.

\bibitem{cpfnet}
S.~Feng, H.~Zhao, F.~Shi, X.~Cheng, M.~Wang, Y.~Ma, D.~Xiang, W.~Zhu, and
  X.~Chen, ``{CPFNet}: Context pyramid fusion network for medical image
  segmentation,'' \emph{IEEE Transactions on Medical Imaging}, vol.~39, no.~10,
  pp. 3008--3018, 2020.

\bibitem{mru}
N.~Ibtehaz and M.~S. Rahman, ``{MultiResUNet}: Rethinking the {U-Net}
  architecture for multimodal biomedical image segmentation,'' \emph{Neural
  Networks}, vol. 121, pp. 74--87, 2020.

\bibitem{att}
O.~Oktay, J.~Schlemper, L.~L. Folgoc, M.~Lee, M.~Heinrich, K.~Misawa, K.~Mori,
  S.~McDonagh, N.~Y. Hammerla, B.~Kainz \emph{et~al.}, ``Attention {U-net}:
  Learning where to look for the pancreas,'' \emph{arXiv preprint
  arXiv:1804.03999}, 2018.

\bibitem{gated}
J.~Schlemper, O.~Oktay, M.~Schaap, M.~Heinrich, B.~Kainz, B.~Glocker, and
  D.~Rueckert, ``Attention gated networks: Learning to leverage salient regions
  in medical images,'' \emph{Medical Image Analysis}, vol.~53, pp. 197--207,
  2019.

\end{thebibliography}

\end{document}